# SS 433 in the crosshairs
## Two recent determinations of the mass ratio

M G Bowler
Department of physics, University of Oxford
Keble Road, Oxford OX1 3RH   (michael.bowler@physics.ox.ac.uk)

**Abstract.**  The unique microquasar SS 433 is a super-Eddington accretor – is the compact object a neutron star, a low mass stellar black hole or a high mass stellar black hole? Knowledge of the ratio $q$ of the masses of the binary components would settle this. I compare the results of two recent and very different determinations of $q$. They are both robust and agree very well; each illuminates the other. The mass of the compact object emerges as $\sim 14\ M_\odot$.

**1.Introduction**  I introduce this informal note with an illustration of the result; Fig.1. The abscissa is the ratio of the mass of the compact object to that of the companion, $q.$ The close intersection of two straight lines and a single curve shows immediately that the value of the ratio $q$ is ~0.7.

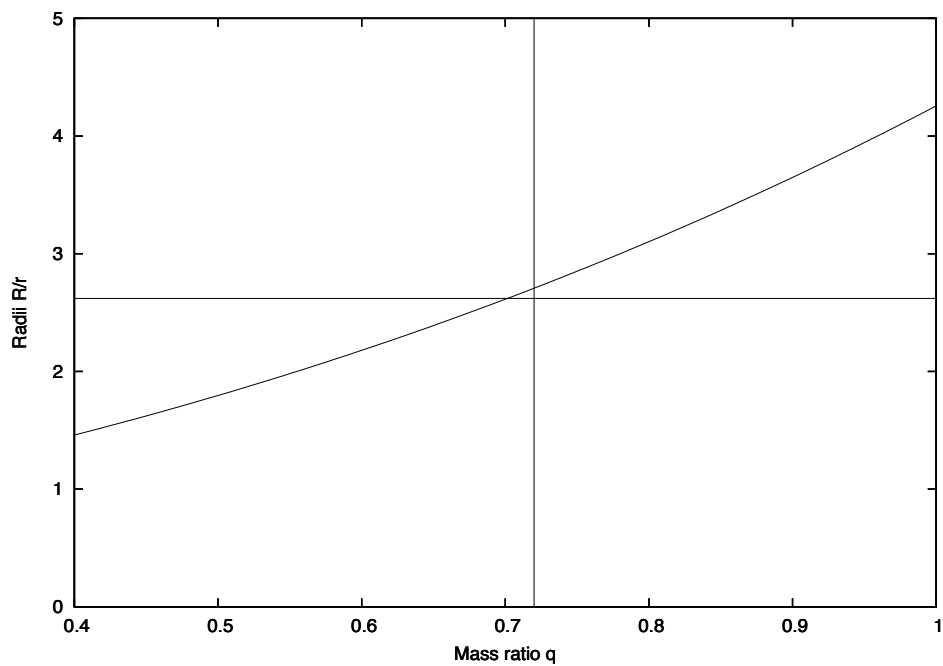

Fig. 1  The mass ratio $q$ determined from the circumbinary disk agrees with the value inferred from constancy of the period, see text.

In Fig.1 the vertical line (or crosshair) passing through $q = 0.72$ is the result of a determination using the remarkable constancy of the 13.08 day period of the binary orbit (Cherepashchuk et al 2018). The ordinate in Fig.1 is the ratio of the inner radius $R$ of the circumbinary disk (Bowler 2011, 2013) to the radius $r$ of the orbit of the compact object. The curve shows the variation of this ratio with $q$, holding constant the known properties of the SS 433 system. The horizontal line (or crosshair) is the value of this ratio obtained from an analysis of the shapes of the He I stationary lines and the variations of these shapes with orbital phase. The intersection gives a second value of $q$. This second method is a new extension of the analyses to be found in (Bowler 2011, 2013).

**2. The binary period.** The period of the binary orbit in SS 433 is 13.08 days and over the last forty years or so has shown no significant change, despite the transfer of $\sim 10^{-4} \, M_\odot \, yr^{-1}$ from the donor to the accreting object. This absence of any significant change has the immediate implication that the mass ratio must be high. It is obvious that were the mass transfer conservative, the constant period would require $q \sim 1$. The transfer is not conservative however, because a strong wind above the accretion disk transports out of the system $\sim 10^{-4} M_\odot \, yr^{-1}$. Nonetheless, if the period does not change the mass ratio must be high. At this opposite extreme where the transferred mass is ejected from the system the minimum value of $q$ is 0.72. For an unchanging period - Fabrika (2004 p 105) quotes $q = 0.7 – 0.8$ from a 1990 calculation. The limits on the accuracy of any such estimate of course depend on the rate of mass transfer and on the limits of the time derivative of the period. In their recent paper Cherepashchuk et al (2018) estimate the value of $q$ using data on the period accumulated over some 28 years, modulus of $\dot{P}/P < 5.7 \, 10^{-7} yr^{-1}$. They find that $q$ is constrained to lie between 0.7 and 0.74. The underlying assumptions are that most of the transferred material is blown from the accretion disk with the specific angular momentum of the accretor and that the rate of mass transfer is $10^{-4} M_\odot \, yr^{-1}$. (Should the mass transfer rate be a factor of ten smaller, the limits expand to 0.59 – 0.94.) The above explains the origin of the vertical line in Fig.1 and this simple study is by itself sufficient to establish the compact object as a stellar black hole of high mass.

**3. He I in the circumbinary disk** The eclipsing system is surrounded by a circumbinary ring or disk that radiates strongly in the visible and infrared. From the separation of the red and blue components of the stationary spectral lines it is found that the glowing material orbits at over 200 $km\ s^{-1}$. The innermost stable circumbinary orbit has a radius ~ 2$A$, where $A$ is the separation of the two components; the inference is that the mass ratio is high, ~0.7 (Blundell, Bowler & Schmidtobreick 2008). The most recent summary of evidence for a circumbinary disk is Bowler (2017). One of the strongest indicators is the shape of the stationary He I line and the ways that shape varies with orbital phase (30 successive days covering over two orbits are shown in Fig. 2 of Schmidtobreick & Blundell 2006). It emerged that He I is stimulated by radiation from the compact object according to the inverse square law (Bowler 2011, 2013). The shapes at various orbital phases are sufficiently well defined to yield the ratio of the radius $R$ of the ring of fire to the radius $r$ of the orbit of the compact object. From this ratio the value of $q$ can be determined. Thus if the radius $R$ were to be 2$A$ and $r$ were to be $A/2$ then the ratio of illumination of the ring closest to the compact object to illumination at the opposite pole would be 25/9, 2.78. Comparing such a model with the data, it is seen that in fact this ratio is close to 5 and hence the ratio $R/r$ is 2.62. Fig.2 shows the configuration of the binary and the circumbinary ring and two panels from Bowler (2011) for data and model at an orbital phase ~0.75, where the compact object is receding fastest. For phase 0.25 the mirror image shape results.

This explains the origin of the horizontal crosshair in Fig.1. There remains the curve relating the ratio $R/r$ to $q$. Given the well established properties of the SS 433 binary, a measurement of $R/r$ is sufficient to determine $q$. Thus if $R/r$ were 4, $q$ would be equal to 1. For the observed ratio of 2.62, we have $q=0.7$, in remarkable agreement with the result from the long term stability of the period of the binary.

The curve in Fig. 1 is given by the relationship

$$\frac{R}{r} = (\frac{v}{V})^2 (1 + q)^3$$

In this note I have assumed the following parameters for SS 433: Binary period 13.08 days, orbital speed ($v$) of compact object 175 $km\ s^{-1}$, orbital speed ($V$) of the ring of fire 240 $km\ s^{-1}$.

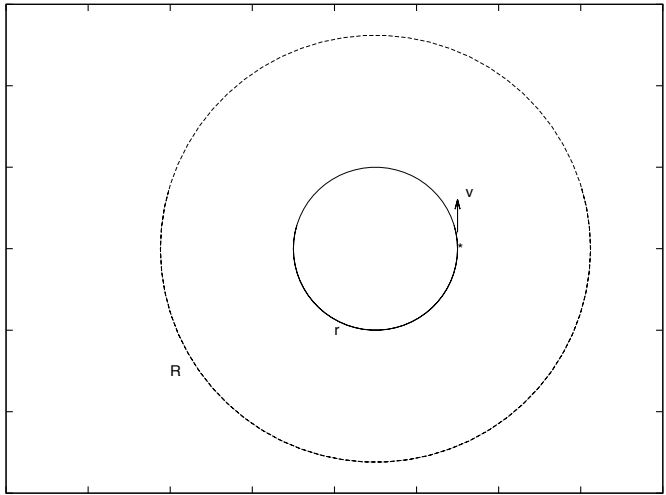

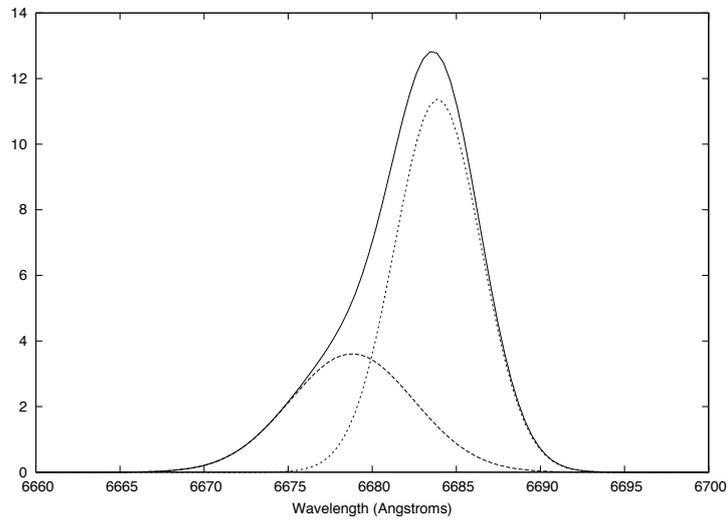

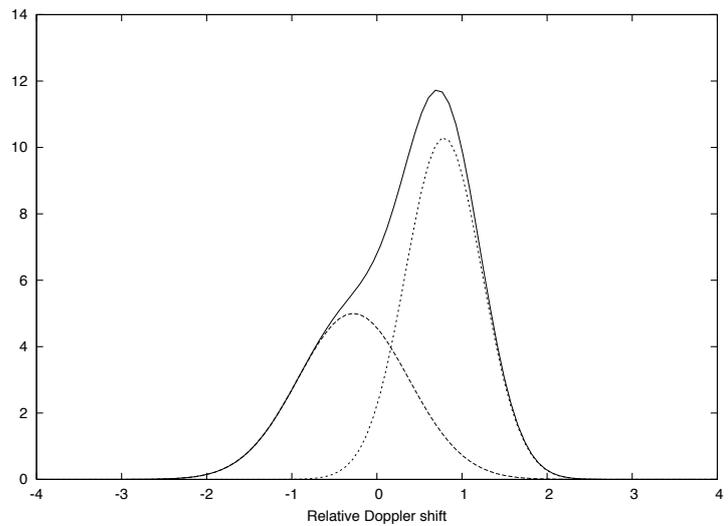

Fig.2 The configuration of the orbits of the compact object (* at radius *r*) and the circumbinary ring and He I line shapes at orbital phase 0.75, see text.

## 4. Discussion

The absence of any detectable change in the binary period over 28 years establishes that $q$ is greater than or equal to 0.7, provided the rate of mass transfer is $\sim 10^{-4} M_\odot \, yr^{-1}$. If most of this mass is ejected from the system with the specific angular momentum of the accretor, then the value of $q$ is within the limits 0.7 – 0.74 (Cherepashchuk et al 2018). Thus even this extremely robust method of determining $q$ is not free from assumptions, but note that the circumbinary torus contains only $\sim 10^{-8} M_\odot$ (Bowler 2013).

The present study of the shape of the stationary emission lines and their variation with orbital phase provides totally independent data. First, the value of the ratio $R/r$ is certainly less than 4 and hence $q <$ 0.9. From Fig.2 (left panel) of Cherepashchuk et al (2018) this implies that less than $\sim 0.1$ of the transfer flux joins the circumbinary disk. At the other end, the lowest value that Cherepashchuk et al (2018) contemplate is 0.6; if this were the case the radius of the ring would be 1.36 $A$ and this is very close to radius of the $L_2$ point, $\sim 1.25\ A$. Finally, the stationary emission lines give the orbital speed of the radiating circumbinary material and the He I line shapes determine the ratio $R/r$. For a fixed value of this ratio, $q$ is a function of that orbital speed. As can be seen from the model calculation in Fig. 2 (bottom panel) the line shape determined by the ratio $R/r$ has no scale, but the orbital speed of the ring is determined by matching the scale of the abscissa to the wavelength scale for the measured line shape. For this same shape, had it turned out that the orbital speed were 200 $km\ s^{-1}$ then $q$ would be 0.51, far below the lowest value permitted by the stability of the orbital period of the binary.

These two recent and independent methods of obtaining the mass ratio lock together, bolstering their assumptions either implicit or explicit.

## 5. Conclusions

The principal conclusion was presented graphically in Fig. 1 in the introduction. It is that the two independent methods of determining $q$ fix its value at $\sim 0.7$ and that subject to the validity of the assumptions used in Cherepashchuk et al (2018) the uncertainty on this value is very small. I shall assume that $q = 0.7 \pm 0.05$. The previously uncertain parameters for the SS 433 binary are then established as below, assuming the other parameters for SS 433 as at the end of section 3:

| | |
|---|---|
| System mass | $35 \pm 3\ M_\odot$ |
| Compact object mass | $14 \pm 2\ M_\odot$ |
| Companion mass | $20 \pm 1\ M_\odot$ |
| Radius $R$ | $1.54 \pm 0.4\ A$ |
| Binary separation $A$ | $5.25 \pm 0.15\ 10^7\ km$ |

It will be noted that $R$, the effective inner radius of the circumbinary disk, is significantly smaller than that of the innermost stable orbit, $\sim 2A$. Were $R$ to be $\sim 2A$ then $q$ would be $\sim 0.9$, but $R/r$ would have to be $\sim 4$. This is ruled out by the He I spectra.

In short, the compact object in the binary system SS 433 is a stellar black hole of high (but not exceptional) mass.

### Acknowledgement

Using the shape of the He I stationary lines to determine the ratio of the radius of the circumbinary ring to the radius of the orbit of the compact object and hence $q$ only occurred to me after I was stimulated by the most interesting paper of Cherepashchuk et al (2018).